# Gradient Optical Diffraction Tomography

Julianna Winnik[1,5,^], Piotr Zdańkowski[1,6,^], Marzena Stefaniuk[2], Azeem Ahmad[3], Chao Zuo[4], Balpreet S. Ahluwalia[3], Maciej Trusiak[1,*]

[1]Warsaw University of Technology, Institute of Micromechanics and Photonics, Boboli 8, 02-525 Warsaw, Poland
[2]Laboratory of Neurobiology, BRAINCITY, Nencki Institute of Experimental Biology of Polish Academy of Sciences, Poland
[3]Department of Physics and Technology, UiT-The Arctic University of Norway, Tromsø, 9037, Norway
[4]Smart Computational Imaging Laboratory (SCILab), School of Electronic and Optical Engineering, Nanjing University of Science and Technology, Nanjing, Jiangsu Province 210094, China
[^]contributed equally as first co-authors
Emails: [5]julianna.winnik@pw.edu.pl, [6]piotr.zdankowski@pw.edu.pl, [*]maciej.trusiak@pw.edu.pl

## Abstract

*Optical diffraction tomography (ODT) enables non-invasive information-rich 3D refractive index (RI) reconstruction of unimpaired transparent biological and technical samples, crucial in biomedical research, optical metrology, materials sciences, and other fields. ODT bypasses the inherent limitations of 2D integrated quantitative phase imaging methods. To increase the signal-to-noise ratio easy-to-implement common-path shearing interferometry setups are successfully combined with low spatiotemporal coherence of illumination. The need for self-interference generated holograms, with small shear values (fitting within the coherence length), critically impedes the analysis of dense and thick samples, e.g., cell cultures, tissue sections, and embryos/organoids. Phase gradient imaging techniques, deployed as a popular solution in the small shear regime, up to now were constrained to 2D integrated quasi-quantitative phase imaging and z-scanning for depth resolution. To fill this significant scientific gap, we propose a novel gradient optical diffraction tomography (GODT) method. The GODT uses coherence-tailored illumination-scanning sequence of phase gradient measurements to tomographically reconstruct, for the first time to the best of our knowledge, a derivative of the 3D RI distribution (in the shear direction) with clearly visible 3D sample structure and high sensitivity to its spatial variations. We present a mathematically rigorous theory based on the first-order Rytov approximation behind the new method, validate it using simulations deploying numerical Shepp-Logan target and corroborate experimentally via successful tomographic imaging of the calibrated nano-printed cell phantom and efficient examination of neural cells. This novel imaging modality opens new possibilities in biomedical quantitative phase imaging, advancing the field and putting forward a first of a kind contrast domain: 3D RI gradient.*

## 1. Introduction

Noninvasive examination of transparent microsamples, vital, e.g., in bioimaging [1] and technical inspection [2], poses an inspiring challenge continuously stimulating the development of optical instrumentation. Quantitative phase imaging (QPI) [3,4] augments the contrast via mapping the optical phase delay introduced by the sample as it locally varies in optical thickness. Its qualitative predecessors, Zernike phase contrast microscopy [5] and Nomarski differential interference contrast microscopy [6] physically transferred, to some extent, phase features into intensity that is easily observable through an eyepiece but did not pinpoint any interpretable values. Full appreciation of a three-dimensional sample is unlocked by optical diffraction tomography (ODT) [7,8] reconstruction, which generates the 3D refractive index (RI) distribution decoupling it from the physical thickness, otherwise mixed in 2D transmission QPI. Well-established 3D QPI framework [9,10] deploys digital holographic microscopy (DHM) [11] principle to first capture the sequence of holographic projections and then process them to extract phase maps and further transform them into 3D RI via tomographic reconstruction. Holotomography [9,10] is thus a flagship representative of computational imaging [12], which numerically bypasses physical limitations. As holograms are generated upon interference [13,14], coherent sources are employed to spatially and temporally secure high contrast superimposition of object and reference beams, which is especially important in popular off-axis DHM [11,15] operating in challenging large optical path difference regime. The off-axis DHM is advantageous as it provides easy access to phase via Fourier transform method [16]. However, high degree of coherence results in notorious speckle noise [17,18] and spurious interferences and can severely degrade the final result of 2D and 3D phase imaging.

To conquer this caveat low coherence sources are used, physically increasing the hologram and phase signal to noise ratio (SNR). Spatial coherence is altered in so-called pseudothermal light sources [19,20] involving a rotating scattering plate, which keeps laser-high temporal coherence. Alternatively, temporal coherence is decreased by selecting the sources with wide spectral band, e.g., superluminescent [21] and regular LEDs [22] or supercontinuum laser with acousto-optic filters [23], while high spatial coherence can be still ensured by fiber coupling or pinhole illumination. Spatio-temporal coherence can be lowered jointly for tailored effects [24]. Due to coherence length shortening the off-axis hologram recording is no longer straightforward [25-27], therefore researchers tend to move to common-path configurations [28-33]. They operate in the regime of small optical path differences and, additionally, provide a unique set of advantages including optomechanical robustness and compactness of the setup, as well as higher measurement accuracy and stability due to mutual compensation of replicated distortions upon interference of beams traveling very similar path. Notably, the common-path holography usually depends on the shearing configuration in which interfering beams are the copies of the object beam that were mutually transversely shifted using, e.g., diffraction gratings or shearing modules. Therefore, common-path holotomography [34-38] building on the mentioned advantageous features and offering high SNR 3D RI imaging, is applicable only to sparse samples as there is a need for sample-free region acting like reference beam upon self-interference hologram generation. Some well-established 2D/3D QPI techniques, e.g., Fourier Ptychographic Microscopy (FPM) [39], Transport of Intensity Equation (TIE) [40], optimization-based inverse scattering problem solving [41] and quantitative Differential Phase Contrast (qDPC) [42,43] skip recording of holograms and numerically solve for phase from fringe-less intensity measurements benefiting from LED-related high SNR and high resolution approaching the incoherent limit. Although very powerful they have characteristic opto-numerical problems in 3D imaging, e.g., numerical solvers needed to access the phase are prone to errors altering the transfer of low spatial frequencies, and in general the 3D setups are experimentally very hard to calibrate (e.g. FPM).

Common-path shearing scenarios naturally promote differentiation operation. Historically, the first gradient based technique for phase imaging was seminal Nomarski's Differential Interference Contrast (DIC) Microscopy. With the advent of computational imaging and new materials the phase gradient concept flourished via, e.g., 2D and 3D programmable qDPC [44-48], Gradient Light Interference Microscopy (GLIM) [49,50], Gradient Retardance Optical Microscopy (GROM) [51], Quadriwave Lateral Shearing Interferometric Microscopy (QWLI) [52,53], Metasurface Enabled Microscopy [54,55], and Oblique Plane Back Illuminated

Microscopy [56,57], to name only several approaches. Up to now gradient based techniques for QPI used some kind of phase integration [49-60], which is susceptible to both experimental and numerical errors with non-uniform transfer functions and, generally, is costly in terms of additional hardware and recording/computation time. Moreover, the 3D imaging modality was introduced by *z*-scanning [44-48,49,50,56,57], which is not a fully rigorous ODT reconstruction and loses some correlations between the sample three-dimensional structuring.

To fill this significant scientific gap, we propose a novel gradient optical diffraction tomography (GODT) method with a unique set of novel features. The GODT uses illumination-scanning sequence of phase gradient measurements to tomographically reconstruct, for the first time to the best of our knowledge, a derivative of the 3D RI distribution (in the shear direction) with clearly visible 3D sample structure and high sensitivity to its spatial (3D) variations. We deploy low temporal and spatial coherence illumination to boost the gradient-phase and thus gradient-RI signal to noise ratio. The novel GODT method operates in the common-path holographic configuration and employs small shear, which promotes the technique towards 3D imaging of challenging thick specimens, which are too dense to operate in total shear regime and, moreover, whose thickness can otherwise easily cause decorrelation of the interfering beams (object and reference). We report that recently a related ODT concept based on QWLI was proposed [61]. However, in this approach the phase gradients need to first undergo problematic integration to retrieve the object phase distributions before standard ODT reconstruction is evaluated. Moreover, the approach suffers from typical QWLI challenges, i.e., slowly varying phase assumption and the need for specialized diffractive optical element. Notably, the proposed GODT principle could be used for tomographic reconstruction from QWLI data.

Our work includes a mathematically rigorous new theory based on the well-established first-order Rytov approximation behind the new method. We also validate it using simulations deploying numerical Shepp-Logan target and corroborate experimentally via successful tomographic imaging of the calibrated nano-printed cell phantom and efficient examination of neural cells. This novel imaging modality opens new possibilities in biomedical quantitative phase imaging, advancing the field and putting forward a first of a kind contrast domain: 3D RI gradient.

## 2. Results and Discussion

### 2.1 Theoretical basis of classical and novel gradient-based tomographic phase imaging

#### 2.1.1 Classical ODT

ODT requires acquisition of multiple complex object waves, which is usually achieved with DHM. Tomographic measurement involves acquisition of data, called projections, for multiple viewing perspectives, which can be realized in various ways. In this work we are focused on the most popular ODT configuration utilizing scanning of illumination, which is the most relevant in the context of biomedical applications [62-66] due to its noninvasive character, ease of use and possibility of fast data acquisition [67-69] enabling tomographic investigation of dynamic biological processes. In this ODT setup, a stationary sample, immersed in a medium with refractive index $n_0$, is successively illuminated with a set of plane waves $u_i = A_i e^{i\varphi_i}$ of various wave vectors $\mathbf{k}_i$=[$k_{ix}$, $k_{iy}$, $k_{iz}$ = ($k_0^2$ - $k_{ix}^2$ - $k_{iy}^2$)$^{1/2}$], where $k_0 = 2\pi n_0/\lambda$ and $\lambda$ is the light wavelength in a vacuum. The reconstructed set of scattered waves $u = Ae^{i\varphi}$ is then processed with a tomographic reconstruction algorithm, which recovers the sample 3D RI distribution $n(x,y,z)$. Due to physical limitations, the captured optical fields $u$ can only have the forward propagating spatial frequency components [$k_x, k_y$]:

$$k_x^2 + k_y^2 < k_{max}^2, \qquad (1)$$

where $k_{max} = k_0$. In the experimental reality, the scope of spatial frequencies is further limited by the numerical aperture (NA) of the optical setup, resulting in $k_{max} = 2\pi NA/\lambda$.

Typically, the tomographic reconstruction is based on the Wolf's generalized projection theorem [70], which states that each scattered field $u$ provides information about the object spatial frequency components that are located on a spherical cap, called Ewald sphere, in a 3D object spectrum. This principle is valid under the first-order Born or the first-order Rytov approximation [71] and applies only to weakly scattering samples. The Rytov approximation is known to be less strict in terms of the weak-scattering requirement [71], thus, we will focus on this ODT variant. The Wolf's theory provides a basis for the most popular algorithm for ODT, called direct inversion (DI) or direct interpolation [72], in which the spectral components of the object waves are directly reposited in the 3D object spectrum via interpolation. The Wolf's theory also underpins the filtered backpropagation algorithm [72], which is the space-domain analog to the frequency-domain based DI and which is a generalization of nondiffractive filtered backprojection to the diffractive case. The tomographic reconstruction process with DI is schematically depicted with Fig. 1.

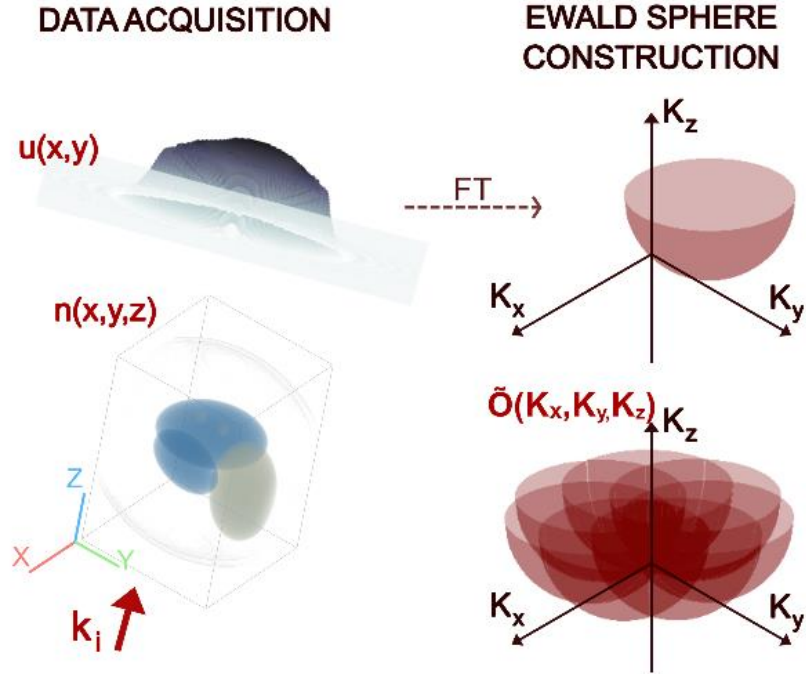

Fig. 1. Illustration of tomographic reconstruction process based on Wolf's theory.

Within the first-order Rytov approximation, the 3D scattering can be expressed as:

$$\tilde{u}_R(K_x, K_y) = \frac{k_0^2}{2ik_z} \tilde{O}(K_x, K_y, K_z). \tag{2}$$

In the applied convention $\sim$ denotes the Fourier transformation and $i$ is an imaginary unit. In the formula, $u_R$ is the Rytov transformed [73] scattered field:

$$u_R = ln\left(\frac{u}{u_i}\right) = ln\left(\frac{A}{A_i}\right) + i(\varphi - \varphi_i). \tag{3}$$

Notice that the Rytov field spectrum $\tilde{u}_R$ (Eq. (2)) is evaluated at the illumination-shifted spectral coordinates:

$$\begin{cases} K_x = k_x - k_{ix} \\ K_y = k_y - k_{iy} \end{cases}, \tag{4}$$

which is related to removal of the phase linear component ($\varphi_i$) in Eq. (3). The coordinates $(K_x, K_y)$ happen to coincide with the transverse Fourier coordinates of the 3D object scattering potential $\tilde{O}$. The axial Fourier component in Eq. (2) is given by:

$$K_z = k_z - k_{iz}, \tag{5}$$

where

$$k_z = \sqrt{k_0^2 - k_x^2 - k_y^2}. \tag{6}$$

It can be notice that Eq. (2) describes the mapping of information contained in the 2D object wave onto the illumination-translated spherical cap (Ewald sphere) in the 3D spectrum of the scattering potential $O$, which is directly related to the RI distribution:

$$n(x, y, z) = n_0\sqrt{1 - O(x, y, z)}. \tag{7}$$

From Eq. 4, it is clear that by varying the illumination direction, it is possible to cover various regions of $\tilde{O}$. Application of a sufficient numbers of diversified viewing perspectives ensures restoration of the sample RI distribution. Notably, in the considered ODT configuration, we deal with the missing frequency problem [74-75], which comes from restriction of the applicable illumination directions range by the NA acceptance angle, resulting in anisotropic (axially limited) 3D resolution of the tomographic imaging.

The outlined Wolf's approach is directly adopted in DI algorithm. The process of filling $\tilde{O}$ is realized with most commonly the nearest neighbor interpolation method. Afterward, $O$ is obtained by the 3D inverse Fourier transformation of $\tilde{O}$ and, finally, $n(x,y,z)$ is evaluated using Eq. (7).

### 2.1.2 Novel GODT

In GODT, the object information is collected with the common-path shearing holography in which the object wave is superimposed with its slightly shifted replica. The interference pattern, captured by the digital sensor, is given by:

$$H(x, y) = |u(x, y) + u(x + \Delta, y)|^2, \tag{8}$$

where $\Delta$ is the shear value. For simplicity, we assume that the shear is purely in $x$ direction; however, our derivation holds for arbitrary transverse shearing direction. Equation (8) can be rearranged to:

$$H = A(x, y)^2 + A(x + \Delta, y)^2 + 2A(x, y)A(x + \Delta, y)\cos[\varphi(x + \Delta, y) - \varphi(x, y)]. \tag{9}$$

It can be noticed that, for sufficiently small $\Delta$, the phase of the registered fringe pattern carries information about the derivative of the object phase in the shearing direction. Specifically, using the limit:

$$\lim_{\Delta \to 0} \frac{\varphi(x+\Delta,y) - \varphi(x,y)}{\Delta} = \frac{d}{dx}\varphi, \tag{10}$$

we can express the phase distribution encoded in the self-interference generated fringe pattern as:

$$\varphi_\Delta(x, y) = \varphi(x + \Delta, y) - \varphi(x, y) = \Delta \frac{d}{dx}\varphi(x, y). \tag{11}$$

Let us now investigate the effect of innovatively applying the shearing holography data as an input to the first-order Rytov-based tomographic reconstruction process. Firstly, in order to simplify the notation, let us skip the step of the object wave correction with the illumination wave, i.e. $u/u_i$, in Eq. (3) and treat $\varphi$ as the object phase with removed illumination-related linear phase component, and $A$ as the object wave amplitude normalized by the illumination amplitude $A_i$. With this simplified notation Eq. (3) reduces to:

$$u_R(x, y) = lnA(x, y) + i\varphi(x, y). \tag{12}$$

The derivative of Eq. (12) in the shearing direction is given by:

$$u_{\Delta R} = \frac{1}{A}\frac{dA}{dx} + i\frac{d\varphi}{dx}. \tag{13}$$

As it can be seen from Eq. (11), the shearing interferometry supplies us with finite element approximation of the derivative of the object phase; however, the derivative of the amplitude is not straightforwardly provided. In some tailored GODT system, it may be possible to separately capture two intensity images of the mutually shifted object beams and in this way compute the finite element approximation of the amplitude derivative:

$$A_\Delta(x,y) = A(x+\Delta,y) - A(x,y) = \Delta \frac{d}{dx} A(x,y). \tag{14}$$

Alternatively, considering that GODT, similarly to ODT, focuses on measurement of transparent weakly-scattering objects, we can assume that the amplitude changes are negligible, thus $A=1$. Then the amplitude terms in Eq. (12) and (13) are set to 0.

Using the shearing holography data, Eq. (13) can be expressed with:

$$u_{\Delta R}(x,y) = \frac{1}{\Delta}\left(\frac{1}{A}A_\Delta(x,y) + i\varphi_\Delta(x,y)\right). \tag{15}$$

Moving forward, by analogy to the conventional ODT (Eq. (2)), in our tomographic reconstruction process, all the Rytov fields (Eq. (15), corresponding to various illumination directions $\mathbf{k_i}$, are Fourier transformed and mapped in the 3D spectral domain, creating $\tilde{O}_\Delta$. Our current goal is to determine what is the information content of the 3D result of this operation. To this aim, let us now focus on the relation between $\tilde{u}_{\Delta R}$ and $\tilde{u}_R$. Using differentiation property of Fourier transform (FT):

$$\frac{d}{dx}f(x) \overset{FT}{\leftrightarrow} iK_x\tilde{f}(K_x), \tag{16}$$

and knowing that $u_{\Delta R}$ is a derivative of $u_R$ we see that $\tilde{u}_{\Delta R}$ and $\tilde{u}_R$ differ just by a differentiation factor $iK_x$:

$$\tilde{u}_{\Delta R} = iK_x \cdot \tilde{u}_R. \tag{17}$$

Let us now apply the same convention as in ODT for mapping of the shearing data in the 3D spectrum (analogy to Eq. (2)):

$$\tilde{u}_{\Delta R}(K_x, K_y) = \frac{k_0^2}{2ik_z}\tilde{O}_\Delta(K_x, K_y, K_z). \tag{18}$$

Combining Eqs (2), (17) and (18) we will obtain the following key relation between ODT and GODT intermediate 3D results:

$$\tilde{O}_\Delta(K_x, K_y, K_z) = iK_x\tilde{O}(K_x, K_y, K_z). \tag{19}$$

Finally, the relation between the scattering potential and the shearing holography data is given by:

$$\tilde{u}_{\Delta R}(K_x, K_y) = \frac{K_x \cdot k_0^2}{2k_z}\tilde{O}(K_x, K_y, K_z). \tag{20}$$

The crucial conclusion from Eq. (19) is that application of the phase derivative data as an input to the first-order Rytov approximation tomographic reconstruction algorithms should not impair this method capability to inverse diffraction and thus generate sharp 3D images with compensated diffraction effects. Moreover, combining Eq. (19) and Eq. (16) we notice that $O_\Delta$ has an elegant analytical interpretation, that it is a derivative of $O$ in the shearing direction. Furthermore, by applying Taylor expansion to Eq. (7) we obtain:

$$n(x,y,z) \approx n_0 - \frac{n_0}{2}O(x,y,z). \tag{21}$$

Basing on this approximately linear relation between $O$ and $n$ we can claim that GODT delivers, for the first time to the best of our knowledge, quantitative 3D information about the derivative of the RI distribution in the shearing direction.

**2.2. Simulation results**

In this section a numerical simulation is used to validate the findings of Sec. 2. Our test object is the modified Shepp-Logan phantom [77] with the real-valued refractive index values ranging from 1.33 to 1.335 and the maximum size of 40 µm. We employed wave propagation method [78-80] to simulate scattering in inhomogeneous medium and generate a set of 180 object waves corresponding to various illumination directions. We applied a conical illumination scenario with a $60^0$ inclination of the illumination direction with respect to the optical axis and the azimuthal step of $2^0$. Other assumed parameters were wavelength of light λ=0.5 µm, refractive index of the immersion liquid $n_0$=1.33, numerical aperture of the imaging system NA=1.33. The size of the simulated object waves was 1000 ×1000 with a sampling interval $dx$=0.1 µm. To emulate the shearing holography data, we numerically shifted the object wave phase and amplitude distributions in $y$ direction by 1 pixel and subtracted them from the original nonshifted versions. We consider two GODT scenarios, first when the finite difference approximation of the amplitude is available, and the second, when the amplitude information is inaccessible, and the uniform amplitude approximation is applied. The three tomographic datasets, corresponding to conventional ODT, GODT with amplitude information, and GODT using phase-only data were submitted to the first-order Rytov approximation tomographic reconstruction with DI algorithm [81]. The results of our test are displayed in Fig. 2 and visualized with Visualization 1.

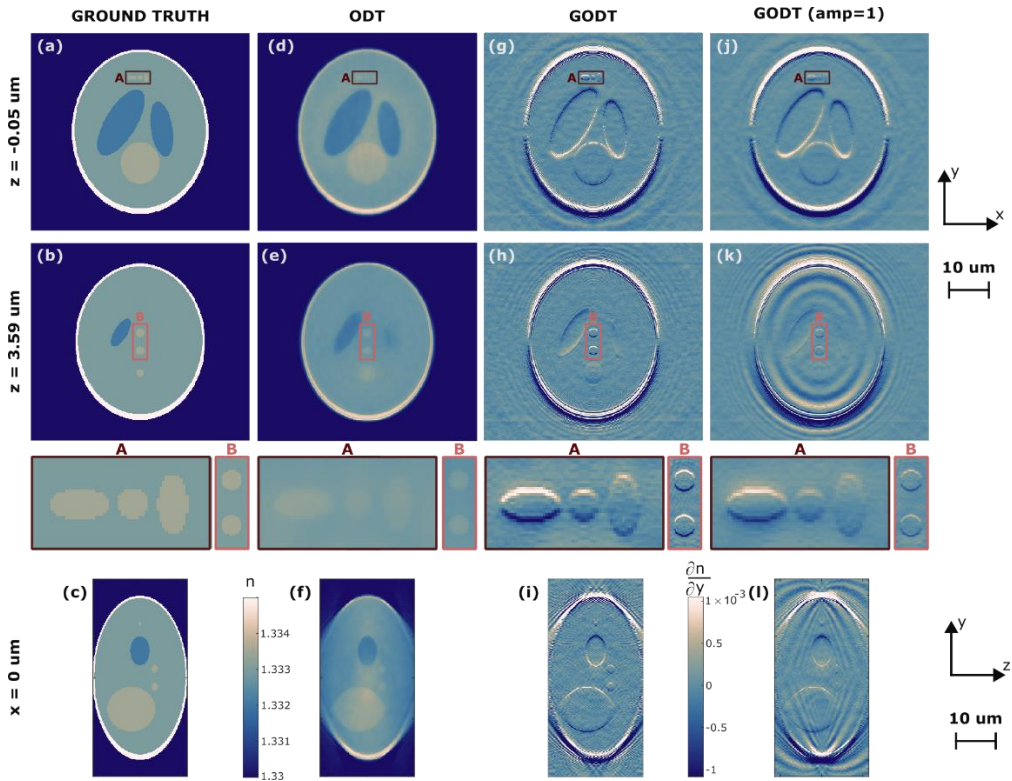

Fig. 2 Numerical test: ground truth (a)-(c) and tomographic reconstructions obtained for various measurement modalities: ODT (d)-(f), GODT (g)-(i), and GODT with uniform amplitude assumption (j)-(l); two transverse cross-sections (two top rows) at different z-depths with zoomed areas of interest A-B and an axial cross-section (lower row). Visualization 1 shows the dynamic z-stacking for each 3D modality and ground truth.

It can be noticed that GODT (Figs. 2(g)-(i)) enabled reconstruction of the sharp 3D object structure with compensated diffraction effects, which validates the tomographic capability of the proposed method. In fact, the obtained GODT result corresponds to derivative of the modelled refractive index distribution in the shearing direction, with the consideration of the limited, anisotropic support of 3D transfer function of the illumination-scanning limited-angle tomographic setup (see section 4.3. for additional corroboration). In our work we consider the primary GODT variant in which a single shearing direction is used and results in different imaging contrast for various orientations of the 3D object features. However, future works may overcome this GODT limitation by combining information obtained for various shear orientations. By analyzing the zoomed areas A and B it can be argued that GODT provides improved visibility with respect to ODT. The GODT is capable of generating high contrast images because it is sensitive to the pace of the RI changes instead of the absolute RI values; additionally, the low coherence of illumination physically "cleans" all recorded holograms from spurious noise. As it can be seen from Figs. 2(k) and 2(l), the lack of amplitude information in GODT leads to diffraction artifacts in the out-of-focus planes $(|z| > 0)$. The strength of this effect depends on how much the amplitudes depart from the uniform distribution, which in turn is determined by the level of scattering nature of the object. The diffraction artefacts related to lack of amplitude information lower the tomographic reconstruction SNR, however, crucially, the structural information content of the reconstruction is basically unchanged and can be readily appreciated.

### 2.3 Experimental validation

### 2.3.1 Nano-printed cell phantom

In order to experimentally validate the proposed GODT method, we used a 3D printed (using two-photon polymerization) cell phantom with preprogrammed RI distribution [82]. The data was captured with a GODT-configured polarization grating aided common-path system [83]. The system uses the set of two polarization diffraction gratings to create self-referencing holograms with flexible adjustment of carrier fringe frequency and the shear value and the orientation between two copies of the object beam. This allows to easily obtain both total-shear and gradient mode interferograms. The details of the GODT experimental system are described in the Methods section.

Figure 3 shows the results of the tomographic reconstruction of the cell phantom using total shear (Fig. 3a) and gradient configuration (Fig. 3b). The results are also visualized with Visualization 2. Firstly, it can be noticed that the GODT tomographic reconstruction process, described in Sec. 2.1, enabled compensating diffraction and retrieving sharp 3D image of the structure of the cell phantom. Moreover, similarly to the numerical simulation results (Sec. 2.2), GODT clearly provides better imaging contrast, and thus sensitivity, for subtle RI changes (see zoomed areas A-D). Cell phantom microstructure is readily appreciable in novel GODT modality with minimized random and background noises. Additionally, it is important to note that the total shear interference setup imposed relatively large optical path difference between interfering beams, enforcing the usage of high coherence light source, i.e., a supercontinuum laser (λ = 633 nm, Δλ = 6 nm). In the gradient configuration, the small shear value allowed application of lowered coherence source, i.e., LED (central wavelength λ = 530 nm Δλ = 35 nm), which improved SNR. Another aspect that is worth to mention, is that the GODT result was obtained using phase information only (i.e., assuming uniform amplitude), which led to residual diffraction fringes in the peripheral areas of reconstructed object (see Sec. 2.2). This effect is visible in Fig. 3(e). Notably, the quality of the GODT reconstruction may be improved with separate acquisition of the amplitude images to construct the amplitude input (Eq. 15) for the GODT tomographic reconstruction algorithm (see Simulation results in section 2.2 and Fig. 2).

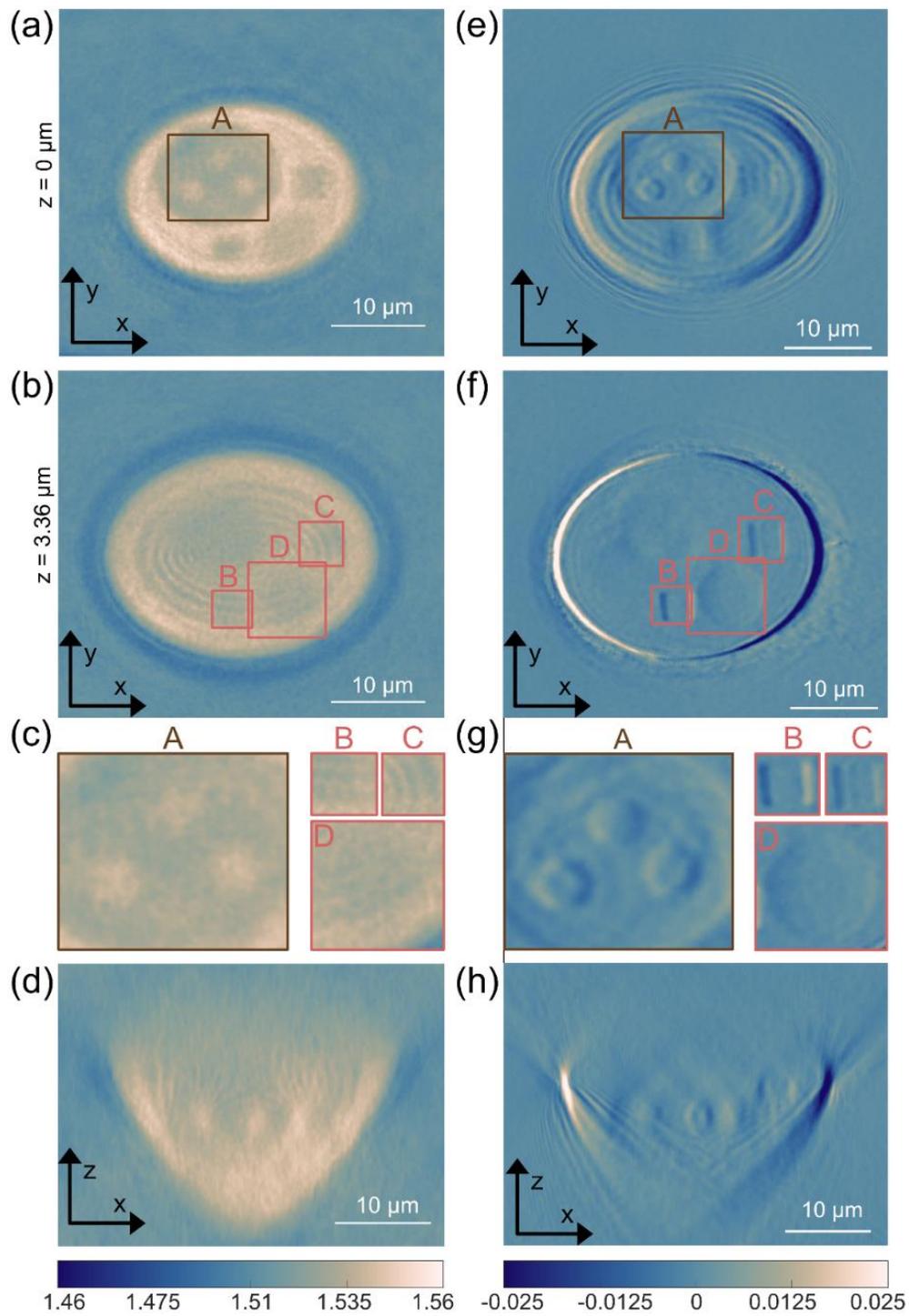

Fig. 3. Tomographic reconstructions of the cell phantom obtained with total-shear ODT (a-d) and GODT configuration (e-h); transverse cross-sections at various depths (a-b), (e-f) with zoomed areas of interest A-D, axial cross-sections (d), (h). Visualization 2 shows dynamic z-stack for total-shear ODT and GODT 3D maps.

### 2.3.2 Neural cells

For the validation of the GODT method on biological samples we used fixed neuronal cells. For this kind of sample that is very confluent, the total-shear imaging mode of our common-path system is not available, as due to the high density of the cells they are constantly overlapping with their replicas. The GODT enables the use of the common path system as the sample confluence is no longer a problem, which constitutes a significant novel value in comparison with other common-path self-referenced setups. Figure 4 shows the result of the GODT reconstruction in the form of two different cross-sections at various depths (Figs. 4(a) and 4(b)) and the 3D view (Fig. 4(c)). The tomographic result is also presented with Visualization 3 and 4. It can be noticed that GODT enabled reconstruction of distinct intracellular features of the neuronal cells at different imaging depths, which proves tomographic capability of the proposed method.

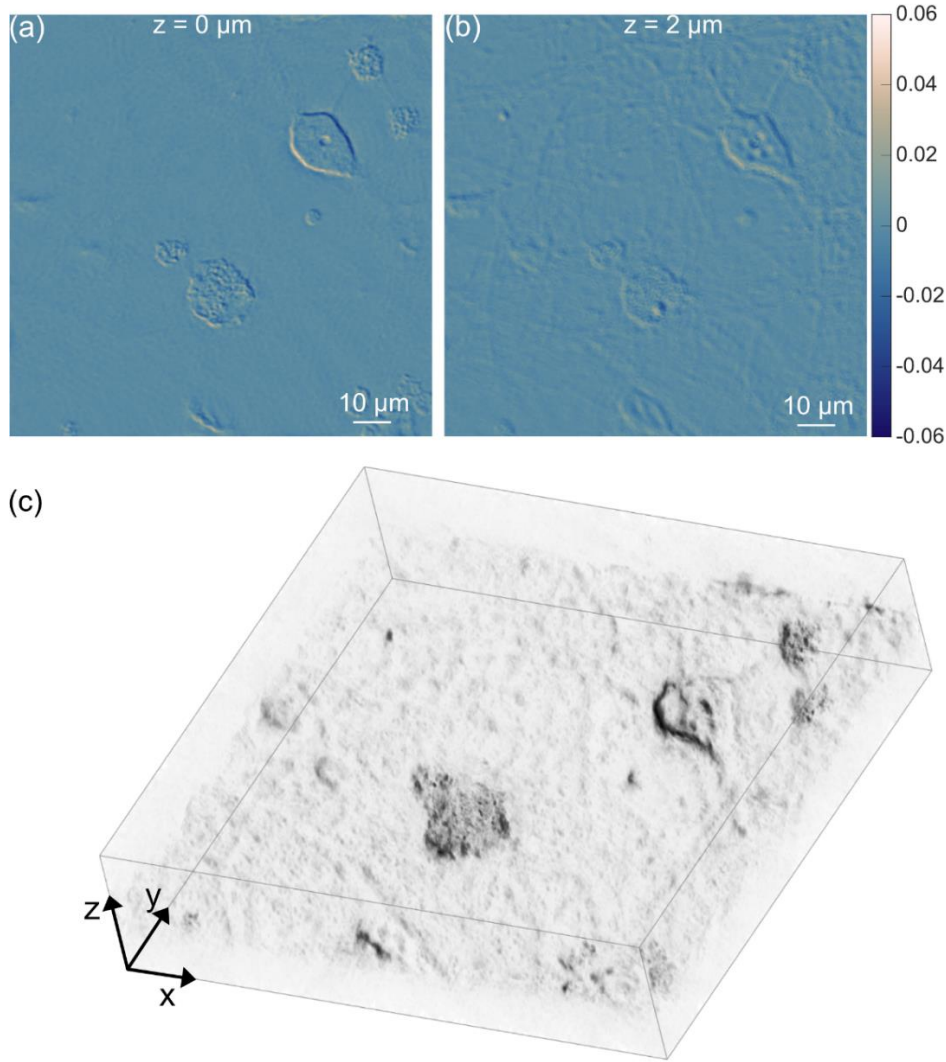

Fig. 4. Results of the GODT reconstruction of the neuronal cells. (a) and (b) show different cross-sections of the reconstructed volume, (c) visualizes the reconstructed 3D volume using maximum intensity projection. Visualization 3 shows dynamic z-stack for 3D neural cells imaged via novel GODT.

### 3. Conclusions

In this work, we have introduced a novel gradient optical diffraction tomography (GODT) method that addresses significant limitations in the existing phase gradient imaging techniques for dense and thick biological samples. By employing a coherence-tailored illumination-scanning sequence of self-referenced commo-path holographic phase gradient measurements, GODT enables first-of-a-kind tomographic reconstruction of the derivative of the three-dimensional refractive index 3D RI distribution along the shear direction. This innovative approach provides high sensitivity to spatial variations and reveals detailed 3D sample structures with unprecedented clarity due to LED illumination and in-line hologram formation architecture.

Our method is grounded in a new fully rigorous theoretical framework based on the first-order Rytov approximation presented in detail in Section 2.1. Validation through simulations (Section 2.2) using the numerical Shepp-Logan phantom, along with experimental demonstrations (Section 2.3) on calibrated nano-printed cell phantoms and neural cells, confirm the efficacy and robustness of GODT. The results showcase the method's capability to capture, via high-quality reconstructions, fine structural details and spatial variations in complex bio-inspired and biological specimens. High-content examination of dense-packed neurons verified GODT biomedically, showing that it is poised to aid cell biology researchers. Due to slowly varying background suppression, which is characteristic to gradient operation, all presented RI distributions have uniquely high signal-to-noise ratio and contrast, innovatively allowing for highly sensitive studies of subtle phase features. A known disadvantage of numerical gradient operation is the amplification of high-frequency noise, however. Nonetheless it is not present in our studies due to low coherence LED illumination deployed.

The putting forward of GODT marks a significant advancement in biomedical quantitative phase imaging by introducing a new contrast domain—the 3D RI gradient. This modality opens new possibilities for imaging practices in cell biology, tissue engineering, and medical diagnostics by providing richer, more detailed information about the internal structures of biological samples without compromising their integrity. Numerical and experimental phantoms corroborated GODT's capability to faithfully image the microstructure of challenging objects. Future research will focus on further refining the GODT technique via amplitude map corrections, exploring its applicability across a broader spectrum of biological specimens, and integrating it with complementary imaging modalities. We anticipate that GODT will become an invaluable tool in the field of optical imaging, facilitating new discoveries and enhancing our understanding of complex biological systems.

## 4. Methods

*4.1 Experimental setup*

The experimental system is shown in Fig. 5. Sample is illuminated with the linearly polarized oblique beam, which then passes through the NA 1.45 100$^\times$ microscope objective (MO) and the tube lens (TL). After the tube lens, the object beam is going through the shearing module (SM). The SM consists of two identical polarization diffraction gratings (PG1 and PG2). First PG1 is diffracting the object beam into two beams by an angle $\theta$, depending on the grating period $\Gamma$: $\theta = sin^{-1}\left(\frac{\lambda}{\Gamma}\right)$, each with opposite handedness of circular polarization (CP). Then, when the CP light is incident on the second PG2 parallel to PG1, it is deflecting each beam by the same angle $\theta$ and inverting their CP, forcing the beams to propagate parallelly. The distance between the gratings $z$ determines the transverse shear $\Delta_M$ between the beams such that $\Delta_M = 2z \cdot \tan\theta$. In the system we used 530 nm LED light source coupled with the multimode fiber with the core diameter of 100 μm. The PG diffraction period Γ = *6.29* μm, resulting in the diffraction angle @ 530 nm $\theta \approx$ *4.83°*. The gratings are separated by $z$ = 0.2 mm, which results in the shear between the beam $\Delta_M$ = *33* μm (which translates to approx. 5 pixels of the CMOS camera). Considering the transverse magnification of the microscope system (100$^\times$), the effective shear value, relevant for Sec. 2.1, is Δ = 0.33 μm).

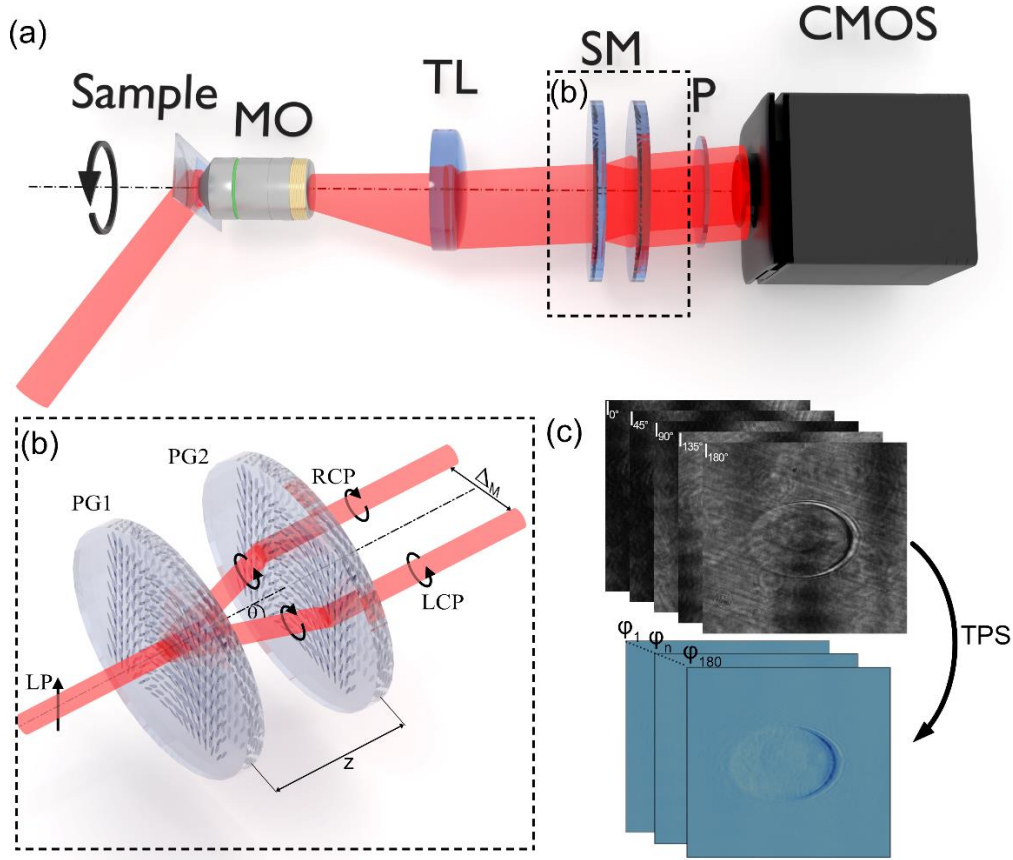

Fig. 5. (a) Layout of the GODT experimental system. MO – microscope objective, TL – tube lens, SM – shearing module, P – polarizer, CMOS – CMOS camera. (b) the shearing module consisting of two polarization gratings PG1 and PG2 separated by the distance *z*, LP – linear polarization, RCP and LCP – righthand and lefthand circular polarizations, respectively. (c) shows the image processing chart, where 5 holograms are acquired for different orientations of the linear polarizer P (0°, 45°, 90°, 135° and 180°), where then phase gradient is calculated using temporal phase shifting (TPS).

For the phase gradient demodulation, we used temporal phase shifting method. The phase shift can be achieved via the rotation of the linear polarizer in front of the CMOS camera, as the two sheared object beams are of opposite handedness of the circular polarization. Thus, the polarizer rotation by α introduces 2α phase shift between the interfering beams.

*4.2 Sample preparation*

The biological samples shown in Fig. 4 were derived from the dissociated hippocampal cultures prepared from postnatal day 0 (P0) Wistar rats [84]. The brains were extracted, and the hippocampi were isolated on ice in dissociation medium (DM; 81.8 mM $Na_2SO_4$, 30 mM $K_2SO_4$, 5.8 mM $MgCl_2$, 0.25 mM $CaCl_2$, 1 mM HEPES [pH 7.4], 20 mM glucose, 1 mM kynurenic acid, and 0.001% phenol red). After rinsing the hippocampi in DM, they were incubated twice in papain solution (100 U in DM) for 15 minutes at 37 °C, followed by three more rinses in DM. The papain activity was neutralized with a trypsin inhibitor dissolved in DM. The hippocampi were then washed three times with plating medium (PM; Minimum Essential Media supplemented with 10% fetal bovine serum and 1% penicillin-streptomycin) and gently triturated until the medium became cloudy. The triturated hippocampi were diluted tenfold in PM and centrifuged for 10 minutes at 1000 × g at room temperature. The resulting cell pellets were resuspended in PM, and the cells were counted and plated at a density of 90,000 cells per 18 mm diameter coverslip pre-coated with 1 mg/ml poly-D-lysine (Sigma-Aldrich) and 2.5 µg/ml laminin (Roche). Two hours after plating, the PM was replaced with maintenance medium (MM; Neurobasal-A with 2% B-27 supplement and 1% penicillin-streptomycin).

*4.3 GODT result equivalence to derivative of the ODT reconstruction*

In Sec. 2.1, basing on the analytical derivation, we argued the GODT provides derivative of the refractive index distribution in the shearing direction. Let us verify this claim with the numerical simulation results from Sec. 2.2. We generated an approximation of the derivative of the ground truth refractive index distribution of the Shepp-Logan phantom using finite difference approach and displayed the results in Fig. 6(a)–(c). The GODT result, presented in Fig. 6(g)-(i), agrees with the ideal result (Fig. 6(a)–(c)) in terms of geometrical structure; however, it is substantially more blurred. That is because the tomographic setup is not capable of transferring the full information about the object due to limited support of the transfer function that is dependent on numerical aperture of the imaging optics, imaging wavelength, and, most importantly, the limited scope of illumination directions. That is exactly the reason why the ground truth result does not fully correspond the ODT result (compare Fig. 2(a)-(c) and Fig. 2 (d)-(f)). As a final test, we numerically evaluate the derivative of the ODT reconstruction from Sec. 2.2. and display it in Fig. 6(d)-(f). The result shows tremendous resemblances to the GODT reconstruction proving the quantitative potential of the proposed method.

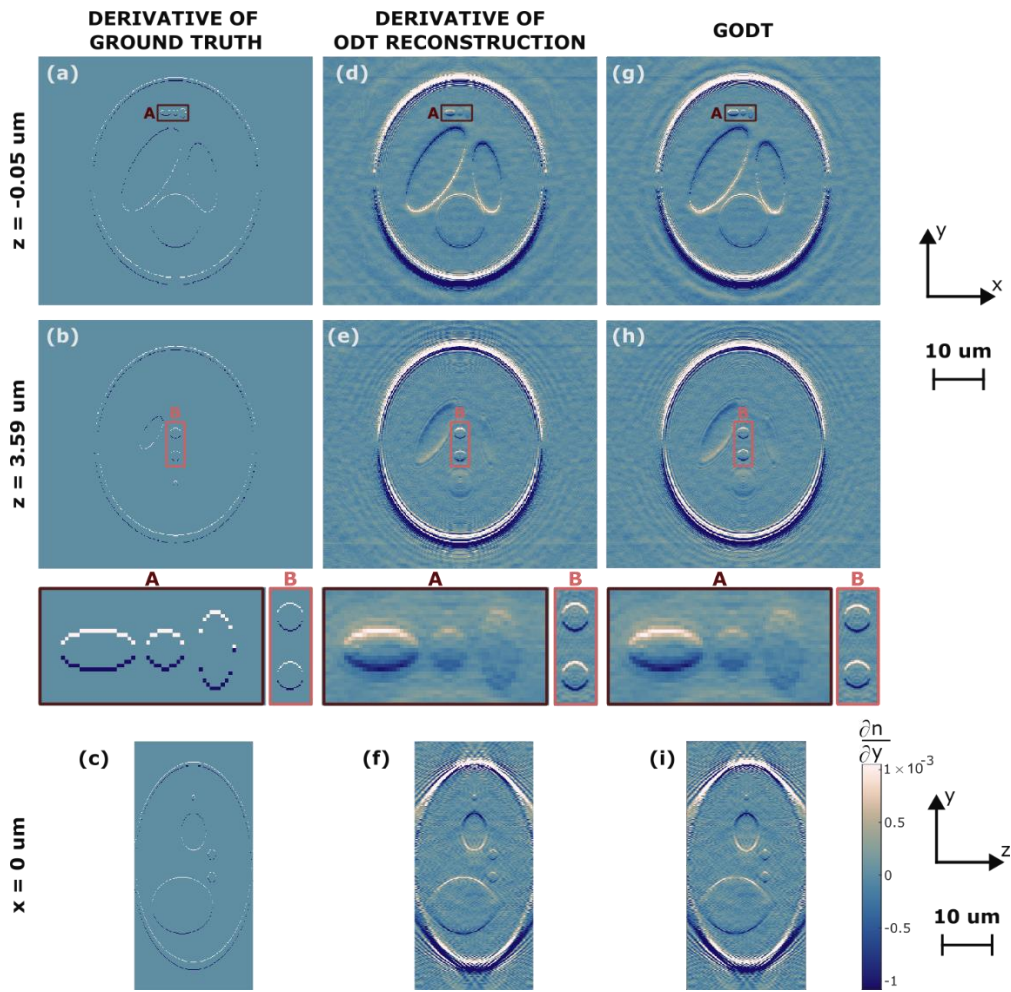

Fig. 6. Numerical test: derivative of ground truth (a)-(c) and ODT result (d)-(f) versus GODT reconstruction (g)-(i); two transverse cross-sections (two top rows) at different z-depths with zoomed areas of interest A-B and an axial cross-section (lower row).

## List of abbreviations

GODT: gradient optical diffraction tomography ODT: optical diffraction tomography QPI: quantitative phase imaging RI: refractive index DHM: digital holographic microscopy SNR: signal to noise ratio FPM: Fourier ptychographic microscopy TIE: transport of intensity equation qDPC: quantitative differential phase contrast DIC: differential interference contrast GLIM: gradient light interference microscopy GROM: gradient retardancy optical microscopy QWLI: quadriwave lateral shearing interferometric microscopy NA: numerical aperture DI: direct inversion FT: Fourier transform MO: microscope objective TL : tube lens SM: shearing module PG: polarization gratings CP: circular polarization TPS: temporal phase shifting LP: linear polarization P0: postnatal day 0 DM: dissociation medium PM: plating medium MM: maintenance medium.

## Availability of data and materials

The datasets used and/or analysed during the current study are available from the corresponding author on reasonable request.

## Competing interests

The authors declare that they have no competing interests.

## Funding

The work was funded by Sheng project: 2023/48/Q/ST7/00172, National Science Center, Poland.

## Authors' contributions

MT and PZ are the authors of the concept of GODT. JW developed a theoretical description of GODT, generated numerical simulation results and performed tomographic reconstructions on the experimental data. MT performed a literature study of existing relevant methods. PZ developed an experimental setup and performed the measurements. JW, MT and PZ were major contributors in writing the manuscript. MS prepared the hippocampal cells for GODT imaging. AA and BSA consulted the methodology and provided important remarks on the draft. All authors read and approved the final manuscript.

## Acknowledgements

We would like to thank Emilia Wdowiak for the cell phantom printing and the Laboratory of Neurobiology from the Nencki Institute for providing neuronal cell cultures.